\documentclass[11pt,superscriptaddress,preprintnumbers, aps,prl,twocolumn, reprint,nofootinbib,longbibliography,tightenlines]{revtex4-2}
\usepackage{amssymb,amsmath}
\usepackage{color}
\usepackage{hyperref}
\usepackage{listings}
\usepackage{physics}
\usepackage{amsmath}
\usepackage{graphicx}
\usepackage{bbold}
\usepackage{qcircuit}
\usepackage{bm}
\usepackage{amsfonts}
\usepackage{comment}
\usepackage{amsmath}
\usepackage{natbib}
\usepackage{multirow}
\usepackage{amsthm}
\usepackage{orcidlink}
\usepackage{forest}
\usepackage[normalem]{ulem}

\usepackage[normalem]{ulem}

\hypersetup{
	unicode=false,          
	pdftoolbar=true,        
	pdfmenubar=true,        
	pdffitwindow=false,     
	pdfstartview={FitH},    
	pdfauthor={},     
	colorlinks=true,       
	linkcolor=blue,          
	citecolor=blue,        
}

\graphicspath{{figures/}}

\definecolor{dkgreen}{rgb}{0,0.6,0}
\definecolor{gray}{rgb}{0.5,0.5,0.5}
\definecolor{mauve}{rgb}{0.58,0,0.82}

\begin{document}
	\preprint{\includegraphics[scale=0.4]{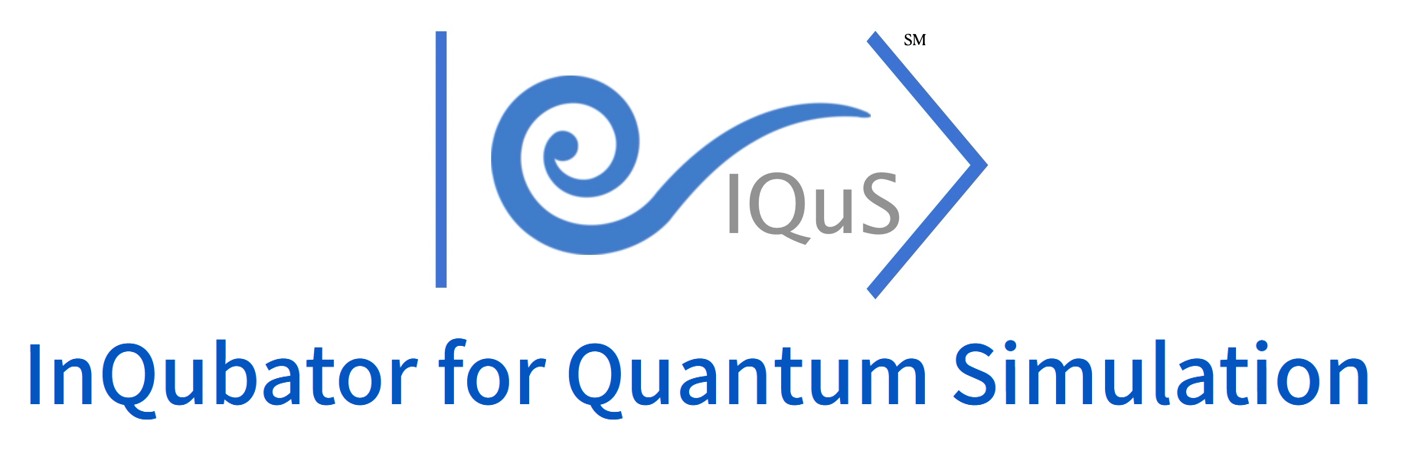} $\qquad\qquad\qquad\qquad\qquad\qquad\qquad$$\qquad\qquad\qquad\qquad\qquad\qquad\qquad$  Preprint number: IQuS@UW-21-054}
	\hspace{0.1cm}
	
	\title{Quantum Imaginary Time Propagation algorithm for preparing thermal states}
	
	\newcommand{\iqusfil}{InQubator for Quantum Simulation (IQuS), Department of Physics,
		University of Washington, Seattle, Washington 98195, USA}

	\author{Francesco~Turro \orcidlink{https://orcid.org/0000-0002-1107-2873}}
	\affiliation{\iqusfil}
	
	\begin{abstract}
		
		Calculations at finite temperatures are fundamental in different scientific fields, from nuclear physics to condensed matter. Evolution in imaginary time is a prominent classical technique for preparing thermal states of quantum systems.
		We propose a new quantum algorithm that prepares thermal states based on the quantum imaginary time propagation method, using a diluted operator with ancilla qubits to overcome the non-unitarity nature of the imaginary time operator. The presented method is the first that allows us to obtain the correct thermal density matrix on a general quantum processor for a generic Hamiltonian. 
		We prove its reliability in the actual quantum hardware computing thermal properties for two and three neutron systems.

	\end{abstract}

	\maketitle
	
	%
	%
	Calculations at finite temperature are essential for understanding quantum systems across scientific fields. In particular, the thermodynamic properties of nuclear matter play a crucial role in heavy-ion collisions, astrophysics, and general nuclear applications. Some examples are nuclear reactions in the evolution of matter in the early universe and inside the core of  stars~\cite{RevModPhys_83_195,annurev-astro-081811-125543,annurev-nucl-020620-063734}, supernova explosions and the phase diagram of QCD~\cite{de2010simulating,shuryak2017strongly}. The recent detection of gravitational waves~\cite{abbott2017gw170817} can provide constraints on the equation of state of nuclear matter at high densities, used, for example, to describe the composition of Neutron Stars~\cite{burgio2021neutron,lattimer2021neutron,haensel2007neutron}.
	
	According to statistical mechanics, the idealization of the density matrix that describes the quantum system with a thermal bath at temperature $T$, and the thermal expectation value of an observable $O$, are given by  
	\begin{equation}
		\rho = e^{-\beta H}\,, \qquad 
		\langle O \rangle = \frac{\Tr{\rho O}}{Z_0} \,, \label{eq:thermal_states}
	\end{equation}
	where $H$ is the Hamiltonian of the system, $\beta=\frac{1}{k_B T}$, $k_B$ the Boltzmann constant, and $Z_0=\Tr{\rho}$ is the partition function.
	
	The imaginary time propagation (ITP) method is a popular classical algorithm to prepare a thermal state of a quantum system. This algorithm was originally designed to prepare the ground state of quantum systems, where one dissipates an arbitrary quantum state to reach the ground state using the ITP operator,  $e^{-\tau\, H}$, where $\tau$ is the imaginary time. Well-known classical techniques for computing thermodynamic properties are Quantum Monte Carlo methods and their improvements, like Auxiliary Field Monte Carlo \cite{baeurle2002field}, Continuous-time Quantum Monte Carlo \cite{gull2011continuous} and Path Integral Monte Carlo \cite{barker1979quantum}. It is a notorious problem that the required classical computational resources grow exponentially with the number of particles. Moreover, many nuclear systems are mainly composed of fermions. Hence, the fermion sign problem emerges, slowing down the progress in studying complex systems~\cite{troyer2005}.
	Following Feynman's idea on the efficiency in simulating quantum systems using quantum hardware~\cite{feynman}, it is desired to develop quantum algorithms that efficiently compute the thermodynamic properties of quantum systems. Different quantum algorithms that implement imaginary time methods and thermal state preparation have been proposed~\cite{motta2020determining,mcardle2019variational,warren2022adaptive,holmes2022quantum,sagastizabal2021variational,consiglio2023variational,turro2022imaginary}. 
	Refs.~\cite{motta2020determining,mcardle2019variational} presents a hybrid quantum-classical algorithm, called Quantum Imaginary Time Evolution (QITE), based on variational ansatz. Applications of the QITE method for thermal states can be found in Ref.~\cite{motta2020determining,davoudi2022toward,sun2021quantum,getelina2023adaptive,wang2023critical,leadbeater2023nonunitary}. Instead, Ref.~\cite{turro2022imaginary} illustrates a quantum algorithm for implementing the imaginary time operator using ancilla qubits with a unitary operator.
	
	This letter presents the first quantum algorithm (to the best of our knowledge) that allows us to obtain the correct density matrix on quantum hardware. This is done by implementing a modified version of the QITP operator of Ref.~\cite{turro2022imaginary}.  
	Moreover, the proposed quantum algorithm is independent of the initial variational ansatz and of the set of classical variables that one has using QITE. In particular, this algorithm can be implemented in studying phase transition, because QITE may become prohibitively expensive due to the large correlation length.

	In this letter, we start from the work of Ref.~\cite{turro2022imaginary}, presenting a quantum algorithm that prepares the thermal state implementing the imaginary time propagation. We also upgrade the QITP algorithm of Ref.~\cite{turro2022imaginary}, improving success probability. We implemented the proposed algorithm in computing the partition function $Z_0$ for spin systems of two and three neutrons in the IBM~\cite{ibm} and Quantinuum H$1$-$1$ quantum hardware~\cite{quantinuum}. We also evaluate the thermal expectation value of some observables. The obtained results are compatible with the analytical behavior. 
	
	%
	%
	\textit{Quantum algorithm. } 
	Our algorithm starts with the initialization of $n_s$ qubits, where we map our physical system, in the so-called maximally mixed state whose density matrix is given by $\mathbb{1}/ 2^{n_s} $ ($\mathbb{1}$ indicates the $2^{n_s}\times 2^{n_s}$ identity matrix). The first proposal was discussed in Ref.~\cite{white2009minimally}, using $n_s$ ancilla qubits (details and our implementation can also be found in App.~A). 
	For small quantum systems, $2-4$ qubits, the ancilla cost is not expensive, but this can be a limitation for bigger systems. However,  this initialization is not strictly necessary, and improvements can be implemented, like minimally entangled typical
	thermal states \cite{white2009minimally,stoudenmire2010minimally} and Canonical Thermal Pure
	Quantum State~\cite{sugiura2013canonical,sugiura2012thermal}. 
	
	%
	%
	After the initialization of the state in the quantum processor,  we should implement the imaginary time operator. We start by summarizing the basic steps of the algorithm in QITP~\cite{turro2022imaginary}.
	Being the imaginary time operator non-unitary, one could work in a diluted Hilbert space to employ a unitary form. Explicitly, we add an ancilla qubit in the $\ket{0}$ state. Then, we implement the following operator
	\begin{equation}
		QITP_{gs} = \begin{pmatrix}
			\frac{e^{-\tau (H-E_T)}}{\sqrt{1+e^{-2\,\tau (H-E_T)}}} & \frac{1}{\sqrt{1+e^{-2\,\tau (H-E_T)}}}\\
			\frac{-1}{\sqrt{1+e^{-2\,\tau (H-E_T)}}} & \frac{e^{-\tau (H-E_T)}}{\sqrt{1+e^{-2\,\tau (H-E_T)}}}\\
		\end{pmatrix} \,,
	\end{equation}
	where $E_T$ indicates the so-called trial energy, an algorithm parameter that should be tuned (see Ref.~\cite{turro2022imaginary} for more details). Then, after the action of the $QITP_{gs}$ operator and measuring the ancilla qubit in $\ket{0}$, the system state is closer to the ground state than the initial one due to the action of $\frac{e^{-\tau (H-E_T)}}{\sqrt{1+e^{-2\,\tau (H-E_T)}}}$ operator.
	
	We have to modify the form of the $QITP_{gs}$ operator such that, after measuring the ancilla qubit $\ket{0}$, we get the correct form of thermal state $e^{-\beta H}$. A straightforward modification is described by
	\begin{equation}
		QITP_{th} = \begin{pmatrix}
			\sqrt{p}\,e^{-\tau (H-E_T)} & \sqrt{ 1- p\,e^{-2\tau (H-E_T)}}\\
			-\sqrt{ 1-p\, e^{-2\tau (H-E_T)}} &\sqrt{p}\,e^{-\tau (H-E_T)}\\
		\end{pmatrix} \,,\label{eq:thermal_QITP_op} 
	\end{equation}    
	where $0<p\le 1$ is a free parameter, representing the success probability in measuring the ancilla qubit in the $\ket{0}$ state in the limit of $\tau \xrightarrow[]{}0 $. For $E_T\le E_0$ ($E_0$ represents the ground energy), we can also set $p=1$, which removes the exponential decay of the success probability. Details for implementing this ITP operator can be found in App.~B. 
	This form requires the least ancilla qubit number (only one) for implementing $QITP_{th}$. Blocking encoding and qubitization~\cite{low2019hamiltonian,low2017optimal,tang2023cs} can be explored to compile the imaginary time operator $e^{-\beta H}$ using more than one ancilla qubit.  
	
	After implementing the operator in Eq.~\eqref{eq:thermal_QITP_op} and measuring the ancilla qubit in the $\ket{0}$ state, we find that the physical qubits are in the correct thermal state (a demonstration can be found in App.~C).
	
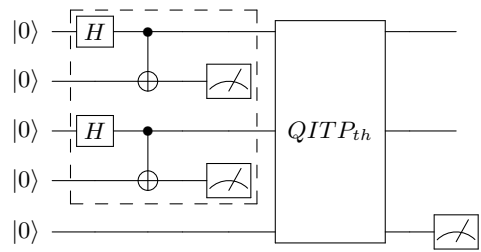
\begin{figure}[t]
		$\Qcircuit @C=1em @R=.7em {
			\lstick{\ket{0}}& \gate{H}& \ctrl{1} &  \qw &\qw &  \multigate{4}{QITP_{th}} & \qw&  \qw& \\
			\lstick{\ket{0}}& \qw & \targ &  \qw & \meter \\
			\lstick{\ket{0}}& \gate{H}& \ctrl{1} &  \qw&\qw& \ghost{QITP_{th}} &  \qw&  \qw&\\
			\lstick{\ket{0}}& \qw & \targ &  \qw & \meter\\
			\lstick{\ket{0}}& \qw & \qw &  \qw &\qw& \ghost{QITP_{th}} &  \qw &\meter \gategroup{1}{2}{4}{5}{0.5em}{--}\\
		}$
		\centering
		\caption{Quantum circuits for preparing the thermal state in quantum processors. The dashed square shows the gates for initializing the system qubits on the maximally entangled mixed state.}
		\label{fig:qc_classical}
	\end{figure}

	The steps of the proposed algorithm to prepare thermal states in quantum processors are as follows:
	\begin{enumerate}
		\item Start with all the qubits in the $\ket{0}$.
		\item Implement the gates of the dashed square in Fig.~\ref{fig:qc_classical} (Its action gives us the physical system qubits in $\mathbb{1}/2^{n_s}$ state).
		\item Employ the the $QITP_{th}$ operator of Eq.~\eqref{eq:thermal_QITP_op} with a additional ancilla qubit, setting $\tau=\frac{\beta}{2}$.
		\item Measure the ancilla qubit in $\ket{0}$.
	\end{enumerate}

	In the worst case scenario, the required number of qubits is $2\, n_s+N_\beta$, where $n_s$ qubits are needed to map the system, $n_s$ ancilla qubits to prepare the maximally mixed state, and $N_\beta$ additional qubits to apply the $QITP_{th}$ operator. Moreover, using more ancilla qubits, one can apply the Trotter decomposition to simplify the compilation of the $QITP_{th}$.
	
	The success probability $P_s$ of the proposed algorithm (equal to the probability to measure all the ancilla qubit in $\ket{0}$) for a perfect noiseless quantum computer is given by 
	\begin{equation}
		P_s=\frac{1}{2^{n_s}} \Tr\left[ p^{N_\beta}\,e^{-\beta (H-E_T)}\right] =\frac{p^{N_\beta}}{2^{n_s}} Z(\beta) \,.
	\end{equation}
	Setting $p=1$, the success probability is proportional to the partition function, $Z(\beta)=\Tr\left[ e^{-\beta (H-E_T)}\right]$. However, in the worst case scenario, for $\beta \xrightarrow[]{}+\infty$, the success probability decays  as $\frac{1}{2^{n_s}}$. This is mostly related to the fact that the state of the quantum system is mostly in the ground state. A solution to this decay could be the application of the amplitude amplification method~\cite{brassard2002quantum} to enhance the success probability.
	
	\begin{figure}
		\centering
		\includegraphics[height=4cm]{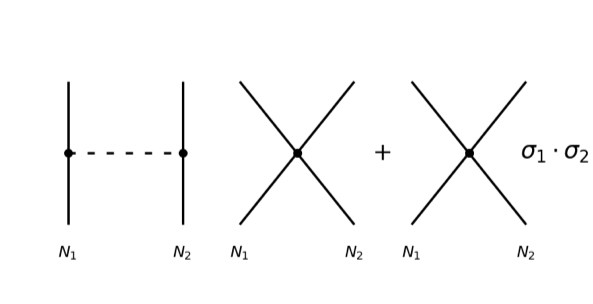}
		\caption{Feynman  diagrams of the Leading Order of Chiral Effective Field Theory}
		\label{fig:feynman}
	\end{figure}
	%
	
	\textit{Results} 
	As a first test, we prepare the thermal state of a spin system of two neutrons fixed in their position. The considered interaction is the spin-dependent part of the leading order of chiral effective field theory~\cite{Chiral_review1,Chiral_review2,tews2016quantum}, using the parameters of Ref.~\cite{holland2020optimal}. The Feynman diagrams of such interaction is shown in Fig.~\ref{fig:feynman}. This system can be mapped to two qubits.

	We implement the thermal $QITP_{th}$ quantum algorithm for different values of $\beta$ using five qubits (two for the system, two for preparing the maximally mixed state and one for implementing $QITP_{th}$). The employed quantum circuits in this work are built with the \texttt{Qiskit} package~\cite{qiskit} and compiled with the \texttt{pytkey} package~\cite{pytket_paper} for the Quantinuum machine. Additionally, we also compute the expectation value of the $\sigma_z$ for the first neutron using the \texttt{H1-1} hardware (adding an extra qubit). Our procedure is described in App.~D. 
	
\begin{figure}[ht] 
		\centering
		\includegraphics[width=\columnwidth]{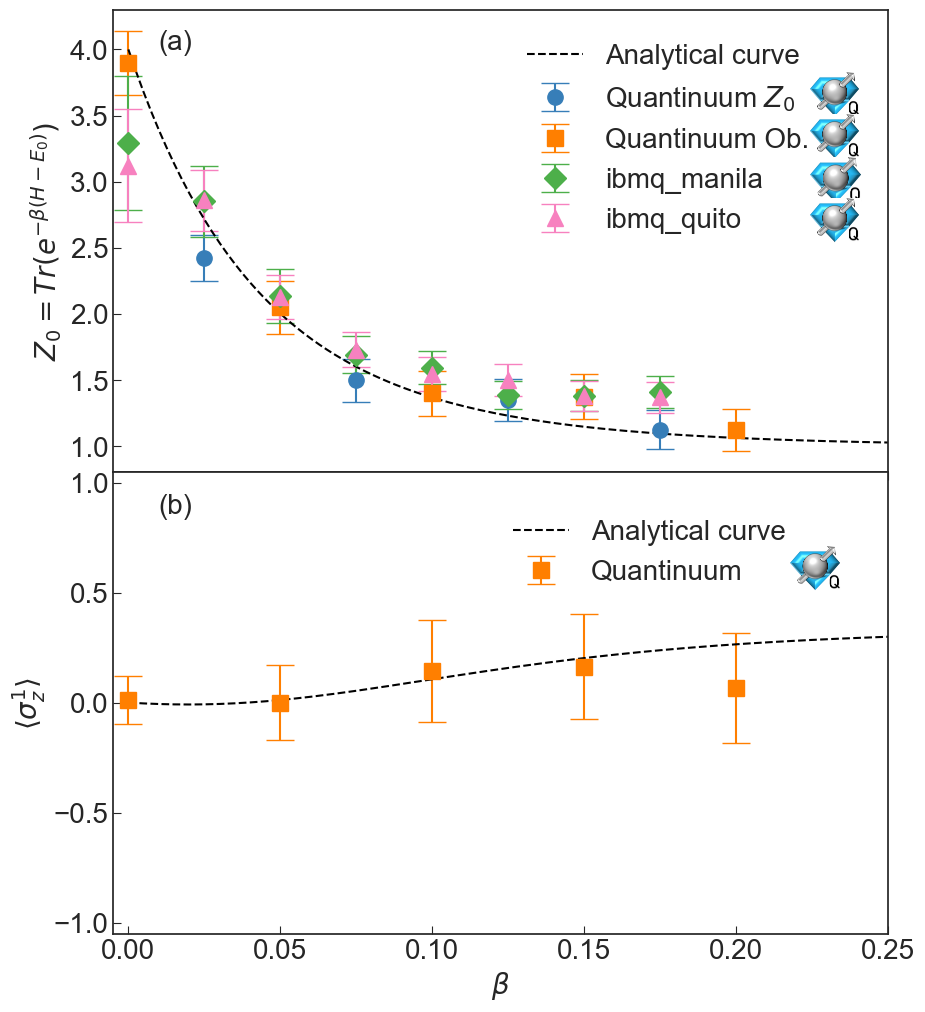}
		\caption{Results for preparing the thermal density matrix for the two neutron spin system. Panel (a) shows the partition function as a function of the inverse of temperature. Panel (b) shows the results of the expectation value of the $\sigma_{z}$ for the first neutron, $\langle \sigma_{z} ^1 \rangle\,=\, \frac{1}{Z_{0}} {\Tr{e^{-\beta (H-E_{T})} \sigma_{z} ^1}}$, as a function of the inverse of temperature.}
		\label{fig:2n_partition_function}
	\end{figure}

	Panel (a) of Fig.~\ref{fig:2n_partition_function} shows our results for the partition function, $Z_0=\Tr{e^{-\beta (H-E_T)}}$. 
	In our tests, we set $E_T$ equal to the ground energy $E_0$ and $p$ in Eq.~\eqref{eq:thermal_QITP_op} equal to $0.8$ to diagnose the algorithm in the non-optimal situation.
	The dashed line represents the analytical values of the partition function (obtained by classically computing the thermal density matrix and tracing it).
	
	The magenta circles and orange squares represent the results obtained from \texttt{H1-1} Quantinuum using both 200 shots. The quantum circuits for the squares compute the partition function and the expectation value, instead, for the circles only the partition function. In the same panel (a), the results from the IBM \texttt{ibmq\textunderscore manila} (green diamonds)  and \texttt{ibmq\textunderscore quito} (pink triangles) are shown as well.  
	While we do not implement any error mitigation methods on the Quantinuuum results, on the IBM hardware we employ the randomizing compiling technique~\cite{wallman2016noise,hashim2021randomized}, using 8 randomized quantum circuits with 64000 total shots (8$\times$8000). Our results are compatible (within two sigma) with the analytical partition function values.
	
	Panel (b) of Fig.~\ref{fig:2n_partition_function} presents the results for the expectation value of $\sigma_z$ (orange squares) using the same obtained probabilities of the panel (a) with different analysis (see App.~D). 
	The dashed line represents the analytical curve. Also, in this case, the obtained results are compatible with the analytical values. However,  with the increase of $\beta$, the error bars get larger due to statistical error because the probability of measuring the ancilla qubits in $\ket{0}$ decreases. A good solution would be to run the quantum circuits with more shots, reducing the statistical errors. Nevertheless, these big error bars are mostly caused by how we compile the hermitian operator. Hence, different strategies, like decomposing in the sum of Pauli matrices, can reduce them. A discussion about tests of these two different compiling methods can be found at the end of App.~D.

	%
	
	%
	
	The next test is preparing thermal states for a spin system of three neutrons. The used Hamiltonian is given by the sum of three two-body terms, ignoring a three-body potential. 
	
	%
	%
	
	We start by evaluating the partition function as a function of $\beta$ implementing a single $\beta$ step. Additionally, we add an extra ancilla qubit to evaluate the expectation value of the Hamiltonian $H$. The final quantum circuits use seven qubits for preparing the thermal state and an additional one to evaluate the expectation value of the Hamiltonian.
	
\begin{figure}[t]
		\centering
		\includegraphics[width=\columnwidth]{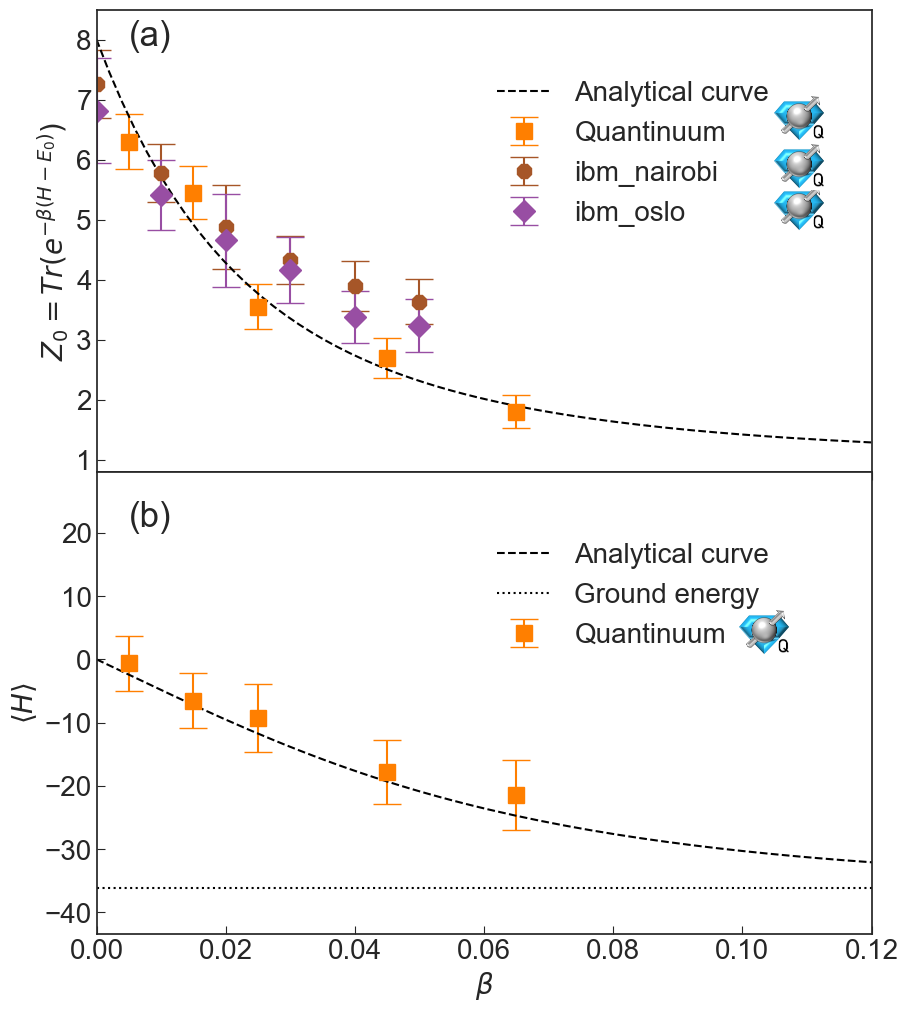}
		\caption{Results of preparing the thermal density matrix for the three neutron spin system. Every point is computed with a single temperature step. Panel (a) shows the partition function as a function of the inverse of temperature. Panel (b) shows the results of obtained expectation value of the Hamiltonian, $\langle H \rangle\,=\, \frac{1}{Z_0} {\Tr{e^{-\beta (H-E_T)} H }}$, as a function of the inverse of temperature. The dashed horizontal line shows the ground energy, which corresponds to the limit for $\beta\xrightarrow[]{}\infty$.}
		\label{fig:3n_single_trotter}
	\end{figure}

	Panel (a) and panel (b) of Fig.~\ref{fig:3n_single_trotter} present the results for the partition function and the expectation value of the Hamiltonian, respectively. The orange squares, brown circles, and violet diamonds indicate the Quantinuum, the IBM \texttt{ibmq\textunderscore nairobi} and \texttt{ibmq\textunderscore olso} results, respectively. The dashed line represents the analytical curves. Even though employing randomizing compiling, we observe the IBM results are noisier than the Quantinuum ones. 
	Indeed, this can be easily explained by the required implementation of swap gates to correctly compile the $QITP_{th}$ operator due to the non-all-to-all connectivity on IBM hardware. The addition of swap gates increases the depth of quantum circuits and the contribution of noise. Nevertheless, our results are still compatible with the analytical values at two sigma.
	One can observe the same increasing behavior of the error bars for the expectation value. Therefore, we can reduce the error bar with a more number of shots.

	%
	%

	%
	%

	\begin{figure}[t]
		\centering
		\includegraphics[width=\columnwidth]{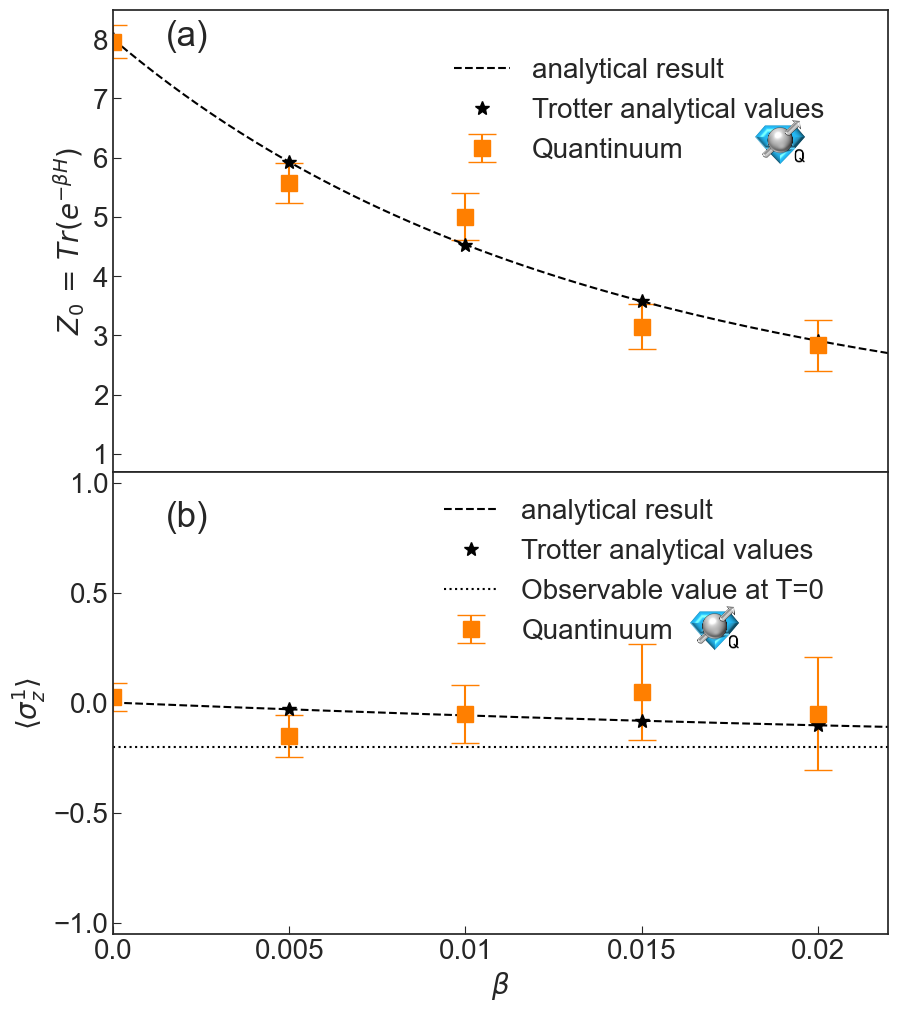}
		\begin{tabular}{|c|cccc|}
			\hline
			$\beta$ & 0.005 & 0.01 & 0.015& 0.02\\
			\hline
			Number of used qubits& 10 & 13 & 16&19\\
			Implemented CNOT gates& 17&29&41&53\\
			\hline
		\end{tabular}
		
		\caption{Results of preparing the thermal density matrix for the three neutron spin system implementing the Trotter decomposition. Panel (a) shows the partition function as a function of the inverse of temperature. Panel (b) shows the results of obtained expectation value of the Hamiltonian, $\langle \sigma^1_z \rangle\,=\, \frac{1}{Z_0} {\Tr{e^{-\beta (H-E_T)} \sigma^1_z }}$, as a function of the inverse of temperature. The dashed horizontal line shows the observable limit for $\beta\xrightarrow[]{}\infty$. The table presents the number of used qubits and CNOT gates.}
		\label{fig:3n_trotterdecomposition}
	\end{figure}
	
	We also test the proposed quantum algorithm implementing the Trotter decomposition for preparing the thermal states of the three neutron spin systems. Specifically, we split $\beta$ in smaller steps and the full thermal propagators into a product of the two-body propagators. Consequently, each layer of the quantum circuit employs a single two-body contribution.
	
	Panel (a) of Fig.~\ref{fig:3n_trotterdecomposition} shows the results obtained from the \texttt{H1-1} Quantinuum processor with orange squares. The dashed line represents the analytical curve applying the full Hamiltonian, and the black stars indicate the analytical values applying the Trotter decomposition. Also, we obtain full compatibility with the analytical values in this case.
	In the runs, we duplicate the number of shots from $200$ to $400$. In the same figure,  we also report the total number of used qubits and implemented CNOT gates for the different $\beta$ values in the lower table. 
	
	We also compute the expectation value of $\sigma_z$ of the first neutron as a function of $\beta$ with the same implemented quantum circuit. as shown in panel (b) of Fig.~\ref{fig:3n_trotterdecomposition}. We observe the same behavior for error bars of Fig.~\ref{fig:2n_partition_function}. A solution may be to increase the number of shots to reduce the error bars.

	%
	%
	\textit{Conclusions }
	This work has presented a quantum algorithm that prepares the quantum thermal states in quantum processors for a generic Hamiltonian. This algorithm is based on the quantum imaginary time propagation method using ancilla qubits to get a unitary version of the imaginary time operator. Upgrading of the quantum imaginary time propagation in Ref.~\cite{turro2022imaginary} has been reported, solving the decaying of the success probability.

	We have discussed the proposed algorithm's validity and reliability in the present quantum hardware. Indeed, we have reported our tests in computing thermal expectation values for simple spin nuclear systems. These simulations have been implemented in different quantum processor hardware, ion-trap (Quantinuum) and superconducting devices (IBM quantum processors). The obtained results are compatible with the analytical values proving the reliability of the proposed quantum algorithm, also in the presence of deep quantum circuits, especially using the \texttt{H1-1} Quantinuum processor. 
	
	This algorithm and the algorithm of Ref.~\cite{motta2020determining} provide a good starting point in preparing thermal states using imaginary time methods on quantum processors. With the leveraging of quantum processors, these algorithms can be applied to solve today's hard problems in nuclear physics, thermalization problems in Quantum Chromodynamics, quantum chemistry, condensed matter applications, and other fields.

	%
	\textit{Acknowledgments} 
	We thank Francesco Pederiva, Alessandro Roggero and the whole IQuS group for useful discussions. In particular, in the IQuS group, we are grateful to Marc Illa Subi\~{n}\`{a}, Anthony Ciavarella, and Martin Savage for the support and correction of the text.
	
	
	This work was supported in part by the U.S. Department of Energy, Office of Science, Office of Nuclear Physics, InQubator for Quantum Simulation (IQuS) (\url{https://iqus.uw.edu}) under Award Number DOE (NP) Award DE-SC0020970 via the program on Quantum Horizons: QIS Research and Innovation for Nuclear Science.
	
	This research used resources of the Oak Ridge Leadership Computing Facility, which is a DOE Office of Science User Facility supported under Contract DE-AC05-00OR22725.
	
	We acknowledge the use of Quantinnum and IBM Quantum services for this work. The views expressed are those of the authors, and do not reflect the official policy or position of IBM or the IBM Quantum team.
	
	A discussion session inspired this work at the "Next-Generation Computing for Low-Energy Nuclear Physics: from Machine Learning to Quantum Computing" IQuS workshop in August 2022.
	
	The obtained results are shown in App.~E. The reported data are obtained from simulations run in February 2023.
	
	\bibliography{references}
	
	\appendix
	
	\section{Appendix A: Preparation of the maximally mixed state}\label{app:Classicalstate_proof}
	
	This section describes a method to obtain the maximally mixed state.
	We start adding ancilla qubits for each system qubit. All the $2\,n_s$ qubits are initialized in $\ket{0}$ state.  Applying the Hadamard gate and CNOT gate, as shown in Fig.~\ref{fig:qc_classical}, and measuring the ancilla qubits, we will obtain the maximally mixed state $\rho=\frac{\mathbb{1}}{2^{n_s}}$.
	
	We will prove this procedure for a single qubit system.  
	After the action of the H and CNOT gates to the state $\ket{00}$, we get
	
	\begin{equation}
		\rho = \text{CNOT}\,H\ket{00}\bra{00}H\,\text{CNOT}= \begin{pmatrix}
			\frac{1}{2} & 0 & 0&  \frac{1}{2} \\
			0 & 0 & 0&0\\
			0 & 0 & 0&0\\
			\frac{1}{2} & 0 & 0&  \frac{1}{2} \\
		\end{pmatrix}\,.
	\end{equation}
	Now, we eliminate the ancilla qubit, for example, measuring it (without interest in the outcome). In mathematical terms, this corresponds to the partial trace. Hence, by doing it, the system state becomes
	\begin{equation}
		\rho^1_{cl}=\Tr_{ancilla}\left[\rho \right]= \begin{pmatrix}
			\frac{1}{2} & 0\\
			0& \frac{1}{2}\\
		\end{pmatrix}\,.
	\end{equation}
	Iterating this method for $n_s$ qubits, we obtain our desired $2^n$-maximally mixed state. Indeed, we have
	\begin{equation}
		\rho_{MME}= \otimes_{n_s} \left(\rho^1_{cl}\right)_{n_s}= \otimes_{n_s} \left(\frac{\mathbb{1_{2\times2}}}{2}  \right)_{n_s}= \frac{1}{2^{n_s}} \mathbb{1}_{2^{n_s}\times2^{n_s}} 
	\end{equation}

	\section{Appendix B: Implementation of $QITP_{th}$ operator}\label{app:implementation_qitp}
	
	This section will discuss how the $QITP_{th}$ operator of Eq.~\eqref{eq:thermal_QITP_op} could be compiled in quantum circuits.
	
	We start by diagonalizing the Hamiltonian $H$,  
	\begin{equation}
		U H U^\dagger = E\,, \label{eq:Hdiagonalization}
	\end{equation}
	where $U$ is the eigenstate matrix and $E$ represent the eigenvalue diagonal matrix.
	
\begin{figure}[th]
		$$\Qcircuit @C=0.5em @R=.5em {
			&\multigate{1}{U} & \qw & \qw& \qw & \qw&  \ctrl{2} & \qw&  \qw& \qw&  \ctrl{2} & \qw& \multigate{1}{U} &\qw\\
			&\ghost{U} & \qw & \qw& \ctrl{1} & \qw&  \qw & \qw&  \ctrl{1}& \qw&  \qw &  \qw & \ghost{U}&\qw\\
			\lstick{\ket{0}}&\qw&\qw &  \gate{R_y}& \targ & \gate{R_y}&  \targ & \gate{R_y}&  \targ& \gate{R_y}&  \targ & \qw &\meter \\
		}$$
		\caption{Example in compiling the $QITP_{th}$ matrix for two qubit system. The upper qubits represent the system ones, and the lower one indicates the ancilla qubit.}
		\label{fig:Greycode}
	\end{figure}
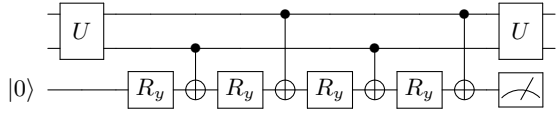

	Therefore, applying Eq.~\eqref{eq:Hdiagonalization} , the $QITP_{th}$ operator can be rewritten in the following form
	\begin{widetext}
		\begin{equation}
			QITP_{th}\,=\,U\, \begin{pmatrix}
				\sqrt{p} e^{-\beta (E-E_T)}&\sqrt{1-p\,e^{-2\beta (H-E_T)}}\\
				-\sqrt{1-p\,e^{-2\beta (H-E_T)}}&\sqrt{p} e^{-\beta (E-E_T)}\\
			\end{pmatrix} \,U^\dagger\,,
		\end{equation}    
	\end{widetext}
	where we assume $E_T\le E_0$.
	The central matrix of the right side of the equation, the matrix with $E$, contains all the physical information. The matrix $U$ is only a change of the computational basis. The central part of $QITP_{th}$ is described by the Cosine Sine decomposition matrix.
	Hence, this can be decomposed as a product of controlled $R_y$ rotations, where the different angle $\theta_i$ are given by $\theta_i=\arccos\left(\sqrt{p} e^{-\beta(E_i-E_T)}\right)$. Moreover, employing the Grey Code \cite{mottonen2005decompositions,barenco1995elementary,tang2023cs}, this operator can be compiled using $2^{n_s}$ CNOT  and  $2^{n_s}-1$ $R_y$ gates. Fig.~\ref{fig:Greycode} shows an example of a two-qubit system.

	\section{Appendix C: Demonstration of thermal preparation}
	
	This appendix proves that we obtain the correct thermal density matrix described by Eq.~\eqref{eq:thermal_states} by implementing all these steps of the proposed quantum algorithm. We assume that we have got the maximally mixed state  $\rho_{MME}=\frac{1}{2^{n_s}} \mathbb{1}$. Applying the $QITP_{th}$ operator and measuring the ancilla in the $\ket{0}$ state we have
	\begin{widetext}
		\begin{equation}
			\begin{split}
				\rho&=P_0\,QITP_{th}\left(\frac{\beta}{2}\right)\,\begin{pmatrix}
					\rho_{MME} & 0\\
					0&0
				\end{pmatrix}  \,\left(QITP_{th}\left(\frac{\beta}{2}\right)\right)^\dagger \,P_0\\
				&= P_0
				\begin{pmatrix}
					e^{-\frac{\beta}{2} (H-E_T)} \rho_{MME} e^{-\frac{\beta}{2} (H-E_T)} & e^{-\frac{\beta}{2} (H-E_T)} \rho_{MME} \sqrt{1-e^{-\beta (H-E_T)}}
					\\
					\sqrt{1-e^{-\beta H}} \rho_{MME} e^{-\frac{\beta}{2} (H-E_T)}             &\sqrt{1-e^{-\beta (H-E_T)}} \rho_{MME} \sqrt{1-e^{-\beta H}}
				\end{pmatrix} P_0\\
				&= \begin{pmatrix}
					\frac{1}{2^n} e^{-\frac{\beta}{2} (H-E_T)} \rho_{MME} e^{-\frac{\beta}{2} (H-E_T)}  &0\\
					0&0\\
				\end{pmatrix} = \frac{1}{2^n} e^{-\beta (H-E_T)} \otimes \ket{0}\bra{0}\,,
			\end{split} \label{eq:rho_th}
		\end{equation}
		
	\end{widetext}
	where $P_0$ indicates the projector to the $\ket{0}$ state of the ancilla. We have proved that the density matrix is proportional to the thermal density matrix.

	\section{Appendix D: Compilation of an observable }
	\label{app:compi_observ}
	This appendix presents how we can compile and evaluate the expectation value of an observable $O$ implementing a single quantum circuit. Usually, the observable is hermitian but not a unitary operator. Therefore, we have to transform it into one.
	
	Using the same spirit of the compilation of the $QITP_{th}$ operator, we define the operator $A$ 
	\begin{equation}
		A= \sqrt{\frac{O-\lambda^O_0}{\left|O-\lambda^O_0\right|}}\,, \label{eq:A_definitation}
	\end{equation}
	where $\lambda_0$ is the lowest eigenvalue of $O$ and $|O-\lambda^O_0|= \max_{\lambda_i} |\lambda_i-\lambda_0|$. With this transformation, we shrink the spectrum of $O$ to be between 0 and 1  (the unitary condition requires this), keeping the hermiticity of $A$. 
	
	We follow the same procedure of the $QITP_{th}$ operator, we add an extra ancilla qubit in $\ket{0}$ state and we define the following unitary operator $U_O$ as
	\begin{equation}
		U_O=\begin{pmatrix}
			A & \sqrt{1-A^2}\\
			-\sqrt{1-A^2} & A\\
		\end{pmatrix}\,.
	\end{equation}
	This new operator is unitary and can be used to evaluate the thermal expectation value of $O$. Indeed, the expectation value of $\langle O \rangle$ is given by
	\begin{equation}
		\langle O \rangle = \langle A^2 \rangle |O-\lambda^O_0| + \lambda_0\,, \label{eq:thermal_O_as_function_A}
	\end{equation}
	where $|O-\lambda^O_0|$ and $\lambda_0$ are defined in Eq.~\eqref{eq:A_definitation}.
	
	To demonstrate the validity of this compiling method, we start by assuming we prepare the system qubits in the maximally mixed state, $\rho_{MME}=\frac{\mathbb{1}}{2^n}$. We add two ancilla qubits, one for the $QITP_{th}$ operator and one for the $U_0$ operator. Applying first the $QITP_{th}$ operator and using the result of Eq.~\eqref{eq:rho_th}, we get:
	\begin{widetext}
		\begin{equation}
			\rho_1=\begin{pmatrix}
				e^{-\frac{\beta}{2} H} \rho_{MME} e^{-\frac{\beta}{2} H} & e^{-\frac{\beta}{2} H} \rho_{MME} \sqrt{1-e^{-\beta H}}& 0 &0\\
				\sqrt{1-e^{-\beta H}} \rho_{MME} e^{-\frac{\beta}{2} H}             &\sqrt{1-e^{-\beta H}} \rho_{MME} \sqrt{1-e^{-\beta H}} & 0 &0\\
				0 &0 &0 &0\\
				0 &0 &0 &0\\
			\end{pmatrix}\,.
		\end{equation}
	\end{widetext}
	
	Then, using $\rho_{MME}=\frac{\mathbb{1}}{2^{n_s}}$ and $A^\dagger=A$, the action of the operator $U_O$ give us 
	
	\newcommand\scalemath[2]{\scalebox{#1}{\mbox{\ensuremath{\displaystyle #2}}}}
	
	\begin{widetext}
		
		\begin{equation}
			\scalemath{0.75}{
				\rho_2=\frac{1}{2^{n_s}} \begin{pmatrix}
					A e^{-\beta H} A & -A e^{-\frac{\beta}{2} H} \sqrt{1-e^{-\beta H}} A  & -A  e^{-\beta H} \sqrt{1-A^2}& A e^{-\frac{\beta}{2} H} \sqrt{1-e^{-\beta H}} \sqrt{1-A^2}\\
					-A \sqrt{1-e^{-\beta H}} e^{-\frac{\beta}{2} H} A& A \sqrt{1-e^{-\beta H}} \sqrt{1-e^{-\beta H}} A & A \sqrt{1-e^{-\beta H}}   e^{-\frac{\beta}{2} H} \sqrt{1-A^2} & -A \sqrt{1-e^{-H \beta}}  \sqrt{1-e^{-H \beta}} \sqrt{1-A^2} \\
					-\sqrt{1-A^2} e^{-\beta H} A & -\sqrt{1-A^2}  e^{-\frac{\beta}{2} H}  \sqrt{1-e^{\beta H}}  A  & \sqrt{1-A^2}  e^{-\beta H}  \sqrt{1-A^2} & -\sqrt{1-A^2} e^{-\frac{\beta}{2} H}   \sqrt{1-e^{-H \beta}} \sqrt{1-A^2} \\
					\sqrt{1-A^2} \sqrt{1-e^{-H \beta}}   e^{-\frac{\beta}{2} H}  A & -\sqrt{1-A^2} \sqrt{1-e^{-H \beta}}   \sqrt{1-e^{-H \beta}}  A & -\sqrt{1-A^2}  \sqrt{1-e^{-H t}}  e^{-\frac{\beta}{2} H} \sqrt{1-A^2}& \sqrt{1-A^2}  \sqrt{1-e^{-H \beta}}  \sqrt{1-e^{-H \beta}} \sqrt{1-A^2}\\  \end{pmatrix}
			}\label{eq:dm_expec_A}\,.
		\end{equation}
	\end{widetext}

	We observe that the probability of measuring both ancilla qubits in $\ket{0}$, indicated with  $P_{00}$,  is equal to the numerator part for the thermal expectation value of $A^2$, $\Tr{A^2\,e^{-\beta H}}$. Moreover, the partition function, $Z_0$, is obtained by the sum of $P_{00}$ and the probability of measuring the ancilla qubit for $U_O$ in $\ket{1}$ and QITP ancilla in $\ket{0}$, indicated with $P_{10}$ (the third diagonal element in Eq.~\eqref{eq:dm_expec_A}). Hence, we have
	\begin{widetext}
		\begin{equation}
			\begin{split}
				P_{00} &= \Tr{A  e^{-\frac{\beta}{2} H} \rho_{MME} e^{-\frac{\beta}{2} H} A^\dagger}= \Tr{ A^2  e^{-\beta H} }\\ 
				P_{00}+P_{10} &= \Tr{A  e^{-\frac{\beta}{2} H} \rho_{MME} e^{-\frac{\beta}{2} H} A} + \Tr{\sqrt{1-A^2}  e^{-\frac{\beta}{2} H} \rho_{MME} e^{-\frac{\beta}{2} H} \sqrt{1-A^2} }\\
				&  = \Tr{e^{-\frac{\beta}{2} H} \rho_{MME} e^{-\frac{\beta}{2} H}}= \Tr{e^{-\beta H}}=Z_0 \,.\label{eq:thermal_expec}
			\end{split}
		\end{equation} 
	\end{widetext}
	where, in the last line, we used the unitary condition of $U_O$, $\Tr{A \rho A + \sqrt{1-A^2} \rho \sqrt{1-A^2}}=\Tr{\rho}$.
	
	Using Eq.~\eqref{eq:thermal_expec}, we can  compute the thermal expectation of $\langle A^2 \rangle$ using 
	\begin{equation}
		\langle A^2 \rangle\,=\,\frac{1}{Z_0} \Tr{ A^2  e^{-\beta H} }\,=\, \frac{P_{00}}{P_{00}+P_{01}} \,.
	\end{equation}
	The expectation value of $O$ is recovered via Eq.~\eqref{eq:thermal_O_as_function_A}.

	\begin{figure*}[t]
		\centering
		\includegraphics[scale=0.38]{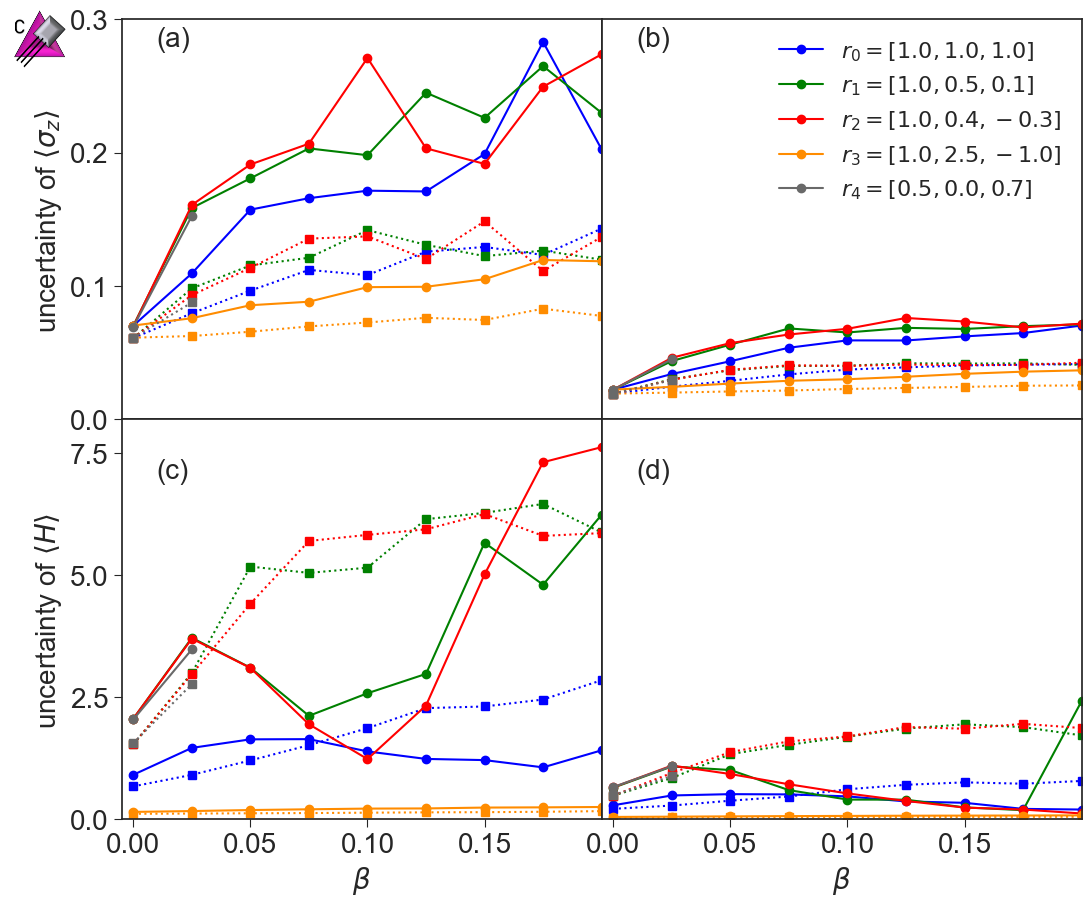}
		\caption{Obtained results for uncertainties of different thermal expectations expanding the observable in Pauli operators (dashed lines) and using the presented method (solid lines) as a function of $\beta$ for different positions of the two neutrons. In (a) and (c) panels, we use 200 shots; in panels (b) and (d), 2000. }
		\label{fig:uncert_pauli_vs_ancilla}
	\end{figure*}
	
	\subsection{Test with the Pauli expansion}
	As reported in the main text, the experimental error bars of thermal expectation values of observable become bigger with the increase of $\beta$. Hence, at the end of this section, we also report our tests about the difference between implementing the proposed algorithm and expanding the observable in the sum of Pauli operators. In particular, we focus on computing the uncertainties of the expectation values of different observables for the spin system of two neutrons by implementing the two different compiling methods. 
	
	We evaluate the uncertainties for a single qubit operator, $\sigma_z$, and a two-qubit operator, the Hamiltonian $H$. They are evaluated as a function of $\beta$ for different nuclear Hamiltonians, where we change the position of the two neutrons. The simulations are implemented on the noiseless IBM emulator.
	
	Panels (a) and (b) of Fig.~\ref{fig:uncert_pauli_vs_ancilla} show the uncertainties for a single Pauli matrix ($\sigma_z$), panels (c) and (d) of Fig.~\ref{fig:uncert_pauli_vs_ancilla} for the Hamiltonian (a generic two-qubit operator). Solid lines indicate the results with the ancilla method, and dashed lines with the Pauli expansion. In panels (a) and (c), we use 200 shots, in panels (b) and (d), 2000 (for the Pauli expansion, we use 200 or 2000 for each quantum circuit).
	
	We observe that the uncertainties for $\sigma_z$ are better for the Pauli expansion. Nevertheless, for a generic single qubit operator, the final uncertainties for the Pauli expansion would generally be twice bigger due to the contribution of the four generators ($\{\mathbb{1},\sigma_x,\sigma_y,\sigma_z\}$). For the two-qubit operator, we notice that the uncertainties of the proposed method (with an extra ancilla)  are similar to or smaller than the results obtained from the Pauli expansion.
	
	Despite the low uncertainties of the Pauli expansion for a single qubit, we have used the presented compiling method to compute the expectation value because we have saved more computational credits.

	\section{Appendix E: Obtained Data}
	The obtained data are shown in the following tables. Tab.~\ref{tab:chi2} reports the reduced chi-square for each simulation.
	
	Tab.~\ref{tab:2n_singlestep} for the two neutrons, Tab.~\ref{tab:3n_singlestep} for the three neutrons with a single step, Tab.~\ref{tab:3n_trotter} for the three neutrons implementing the Trotter decomposition.
	
	\begin{table*}[h]
		\centering
		\begin{tabular}{|c|c|c|c|c|c|c|c|c|}
			\hline
			\multirow{2}{*}{$\frac{\chi^2}{N}$}& \multicolumn{4}{|c|}{2 neutrons} &\multicolumn{3}{|c|}{3 neutrons} &  
			Trotter 3 neutrons  \\
			& \texttt{H1-1} (only $Z_0$)& \texttt{H1-1} (both $Z_0$ and $\sigma_z$)& \texttt{ibmq\textunderscore quito}&\texttt{ibmq\textunderscore manila}& \texttt{H1-1}&\texttt{ibmq\textunderscore olso}&\texttt{ibmq\textunderscore naroibi} & \texttt{H1-1} \\
			\hline
			$Z_0$ & 0.93&0.44&2.2 &1.42&0.62 &0.72 &2.06&0.77\\
			$Ob$ & &0.34&&&0.17&&&0.44\\
			\hline
		\end{tabular}
		\caption{Reduced $\chi^2$ for each simulations}
		\label{tab:chi2}
	\end{table*}
	
	\begin{table*}[h]
		\centering
		\begin{tabular}{|c|cccc|cc|}
			\hline
			\multirow{2}{*}{$\beta$} & \multicolumn{4}{|c|}{$Z_0$} &\multicolumn{2}{c|}{$\langle \sigma_z^1\rangle$ } \\
			& \texttt{H1-1}  &\texttt{ibm\textunderscore manila} &\texttt{ibmq\textunderscore quito} & analytical  & \texttt{H1-1} & analytical  \\
			\hline
			0.025 & 2.4(2) & &2.7& & &\\
			0.075 & 1.50(16) & &1.6& & &\\
			0.125 & 1.35(16) & &1.2& & &\\
			0.175 & 1.12(15) & &1.1& & &\\
			\hline
			0.00 & 3.9(2) & & & 4.0& 0.01(11) &0.00\\
			0.05 & 2.1(2) & & & 2.0& 0.00(17) &0.01\\
			0.10 & 1.40(17) & & & 1.4& 0.1(2) &0.11\\
			0.15 & 1.38(17) & & & 1.1& 0.2(2) &0.20\\
			0.20 & 1.12(16) & & & 1.1& 0.1(3) &0.27\\
			\hline
			0.00 & & 3.3(5) & 3.1(4) & 4.0& &\\
			0.025 & & 2.9(3) & 2.9(2) & 2.7& &\\
			0.05 & & 2.1(2) & 2.1(2) & 2.0& &\\
			0.075 & & 1.69(14) & 1.73(13) & 1.60& &\\
			0.10 & & 1.60(13) & 1.55(13) & 1.37& &\\
			0.125 & & 1.39(11) & 1.50(12) & 1.23& &\\
			\hline
		\end{tabular}
		\caption{Results of the two neutron spin system}
		\label{tab:2n_singlestep}
	\end{table*}
	
	\begin{table*}[h]
		\centering
		\begin{tabular}{|c|cccc|cc|}
			
			\hline
			\multirow{2}{*}{$\beta$} & \multicolumn{4}{|c|}{$Z_0$} &\multicolumn{2}{c|}{$\langle H \rangle$ } \\
			& \texttt{H1-1}  &\texttt{ibmq\textunderscore olso} & \texttt{ibmq\textunderscore nairobi} & analytical & \texttt{H1-1} & analytical  \\
			\hline
			0.005 & 6.3(5) & & & 6.7& 0.0(4) &-2.4\\
			0.015 & 5.4(4) & & & 4.9& -6(4) &-7.2\\
			0.025 & 3.5(4) & & & 3.8& -9(5) &-11.7\\
			0.045 & 2.7(3) & & & 2.5& -17(5) &-19.3\\
			0.065 & 1.8(3) & & & 1.9& -21(5) &-24.7\\
			\hline
			0.00 & & 6.8(9) & 7.3(6) & 8.0& &\\
			0.01 & & 5.4(6) & 5.8(5) & 5.7& &\\
			0.02 & & 4.7(8) & 4.9(7) & 4.3& &\\
			0.03 & & 4.2(6) & 4.3(4) & 3.4& &\\
			0.04 & & 3.4(4) & 3.9(4) & 2.7& &\\
			0.05 & & 3.2(4) & 3.6(4) & 2.3& &\\
			\hline
		\end{tabular}
		\caption{Results of the three neutron spin system without applying the Trotter decomposition}
		\label{tab:3n_singlestep}
	\end{table*}
	\begin{table*}[h]
		\centering
		\begin{tabular}{|c|cc|cc|}
			\hline
			\multirow{2}{*}{$\beta$} &\multicolumn{2}{|c|}{$Z_0$} &\multicolumn{2}{|c|}{$\langle \sigma_z^1\rangle$ }\\
			&\texttt{H1-1}  & analytical &\texttt{H1-1}  & analytical\\
			
			\hline
			0.000& 8.0(3)&8.0&0.03(6)&0.00\\
			0.005& 5.6(3)&5.9&-0.15(10)&-0.03\\
			0.01& 5.0(4)&4.5&-0.05(13)&-0.06\\
			0.015& 3.1(4)&3.6&0.0(2)&-0.08\\
			0.02& 2.8(4)&2.9&0.0(3)&-0.10\\
			\hline
		\end{tabular}
		\caption{Results of the three neutron spin system without using the Trotter decomposition}
		\label{tab:3n_trotter}
	\end{table*}

\end{document}